\documentstyle[12pt]{article}
\newcommand{\prt}{\partial}
\newcommand{\bi}{\bar I}
\begin{document}
\renewcommand{\theequation}{\thesection.\arabic{equation}}
\newcommand{\beq}{\begin{equation}}
\newcommand{\eeq}{\end{equation}}

\title{Instanton Dynamics in the Broken Phase of the Topological
Sigma Model}
\author{A.V.Yung \thanks{Permanent address: Petersburg
 Nuclear Physics Institute
Gatchina, St.Petersburg 188350, Russia}\\
University of Wales, Swansea SA2 8PP, UK}
\date{February 1995}
\maketitle

\begin{abstract}
The topological $\sigma$ model with the black hole metric of the
target space is considered. It has been shown before that this model
is in the phase with BRST-symmetry broken. In particular, vacuum
energy is non-\-zero and correlation functions of observables show
the coordinate dependence. However these quantities turned out to be
infrared (IR) divergent. It is shown here that IR divergences
disappear after the sum over an arbitrary number of additional
instanton-\-anti-\-instanton pairs is performed. The model appears to
be equivalent to Coulomb gas/Sine Gordon system.
\end{abstract}
\vspace*{2cm}
SWAT 95/62\\
hep-th 9502149
\newpage

\section{Introduction}

Considerable progress has been made during last years in the study
of topological field theories (TFT) \cite{1,2}. The main avenue of
these studies is the relationship between $2D$ TFT's and the low
dimensional string theory. It was shown \cite{3,4} that string theory
at $c<1$ equivalent to topological minimal matter \cite{5} coupled to
topological gravity \cite{3}. An important step forward was made in
ref.\cite{6} where the equivalence of $c=1$ string and topological
version of $SL(2R)/U(1)$ $WZW$ coset model \cite{7} at level $k=3$
was shown.

Though much work has been done along these lines one of the most
crucial problem about TFT still remains
unsolved. In the field theory framework TFT has no physical
degrees of freedom: all correlation
functions of observables are just numbers. Therefore, as it was
proposed already in the original Witten's papers \cite{1,2}, we need
some mechanism of the spontaneous breakdown of BRST symmetry for TFT
to have something to do with physics. The idea is that the physical
theory may correspond to the broken phase of the TFT. Then we could
have advantages from the existence of the underlying BRST-\-symmetry
(say, good UV properties) in a theory with some physical degrees of
freedom "liberated".

The above problem persists also in the framework of the topological
string. The $c<1$ non-\-critical string has no physical degrees of
freedom, therefore it is not  surprising that it is actually
topological. The problem of relevance of the string theory beyond the
$c=1$ barrier to any TFT remains unsolved.

In our previous paper \cite{8} the $2D$ topological $\sigma$ model
\cite{2}  with black hole metric \cite{9} of the target space in two
dimensions was considered. Although the target space is not
compact (it has the form of a semi-\-infinite cigar) the model is
shown to possess world-\-sheet instantons. In fact, cigar-\-like
metric appears to be on the ``borderline'' between compact and
noncompact cases and needs careful regularization. The result in
\cite{8} is that the topological version  of the model does have
unsuppressed instantons. The noncompactness of the moduli space of
these instantons produces new divergences. These give rise to the
nonzero vacuum energy and to the coordinate dependence of correlation
functions of observables. Hence, the BRST symmetry is broken
\cite{8}.

Divergences of the integrals over the moduli space of
instanton studied
in \cite{8} are both of UV and IR nature. The UV ones introduce the
UV cutoff parameter $a$ (lattice spacing) dependence of observables.
These are signals of the presence of an extra conformal anomaly in the
theory associated with the noncompactness of the target space.
However the IR divergences are obviously artificial. They should
disappear when all IR singular effects are taken into account.

In this paper we continue to study the topological $\sigma$ model
which has  the target space with the geometry of the
 two-\-dimensional black hole. On one
hand this model can be viewed as a toy example to study the mechanism
of BRST-\-symmetry breaking. On the other hand, the model has the same
geometry of the target space as $SL(2,R)/U(1)$ coset \cite{9}
(although it is not identical to the latter one). This means (in view of
the results in ref.\cite{6}) that the BRST symmetry breaking in the
topological black hole could have some parallel in the $c=1$ string
theory.

Our aim in this paper is to study the physics which emerges in the
broken phase of TFT. In particular, we consider the partition
function and correlation functions of observables in the instanton
vacuum with arbitrary number of instanton--\-anti-\-instanton ($I\bar
I)$ pairs added. We sum over all these IR troublesome effects
 constructing  the effective Lagrangian of the model. After
that IR divergences disappear and the mass scale is dynamically
generated. The model turns out to be equivalent to Sine Gordon (SG)
theory. In fact, the instanton physics in the topological black hole
model appears to be very similar to that in $O(3)$ $\sigma$ model.
 In the
latter model instanton vacuum has the analogous Coulomb gas
description \cite{10}.

The organization of this paper is as follows. In Sec.2 we review the
properties of the
topological $\sigma$ model with black hole metric and show how the
breakdown of the BRST-\-symmetry occurs. In Sec.3 we develop a
certain approximate scheme $(I\bar I$ approximation) and show that
instanton vacuum of the model is equivalent to Coulomb gas/Sine
Gordon (CG/SG) system (or to free massive fermions) in this
approximation. In Sec.4 we calculate the partition function in the
background of two $I\bar I$ pairs and show that (with the proper
definition of the geometry of the modular space of instantons) $I\bar
I$ approximation becomes exact. In Sec.5 we present our final result
for the vacuum energy which turns out to be nonzero and IR-\-finite.
Then in Sec.6 we calculate the two-\-point correlation function of
operators from the cohomology of observables. It appears to be also
IR finite and coordinate dependent. In particular, it shows the power
fall-off at large distances. We interpret this behaviour as a
propagation of the goldstino fermion associated with the broken BRST
symmetry. Sec.7 contains our final discussion.

\section{$d=2$ topological $\sigma$-model with the black hole metric}

First in this section we review some general properties of the
topological $\sigma$ model \cite{2,3} (for a review see also
\cite{11}). The action of the model on the $d=2$ Kahler manifold
reads \cite{2}
\begin{eqnarray}
S&=&\frac{r^2}\pi \int d^2x\left\{\frac 12 g_{ij}(w)\prt_\mu
w^i\prt_\mu w^j -\frac 12 J_{ij}(w)\epsilon_{\mu\nu}\prt_\mu
w^i\prt_\nu w^j\right. \nonumber \\
&-& \left. ig_{ij}\lambda^{\mu i} D_\mu\chi^j -\frac 18 R_{ijk\ell}
\chi^i\chi^j\lambda^{\mu k}\lambda^{\mu\ell} \right\}.
\end{eqnarray}
Here $\mu,\nu=1,2$ are world sheet indices, while $i,j=1,2$ are
target space ones. The world sheet is considered to be flat for
simplicity, while $g_{ij}$ and $J_{ij}$ denote metric and complex
structure of the target space. In this paper we consider target space
metric and complex structure of the form
\begin{eqnarray}
g_{ij}&=&g(w) \delta_{ij} \nonumber \\
J_{ij}&=&g(w)\epsilon_{ij}.
\end{eqnarray}
$D_\mu$ is the covariant derivative, $D_\mu A^i = \prt_\mu A^i
+\prt_\mu w^k\Gamma^i_{k\ell}A^\ell,$ while $\Gamma^i_{kl}$ and
$R_{ijk\ell}$ are connection and
curvature tensor
respectively. Fermion system $\chi^i, \lambda^{\mu i}$ has spins
0, 1 and satisfies the constraint
\beq
\lambda^{\mu i} +\epsilon^\mu_\nu J^i_j \lambda^{\nu j} =0.
\eeq
Fermions play the role of ghosts which cancel out boson degrees of
freedom in correlation functions of observables.

The BRST operator acts as follows
\begin{eqnarray}
\{Q,w^i\}& = & \chi^i, \nonumber \\
\{ Q,\chi^i\}&=&0 \\
\{Q,\lambda^{\mu i}\}&=& 2i(\prt^\mu w^i-\epsilon^{\mu\nu}
J^i_j\prt_\nu w^j)+\lambda^{\mu j}\Gamma^i_{jk}\chi^k. \nonumber
\end{eqnarray}
Observables $O$ of TFT are elements of the $Q$-cohomology
\beq
\{ Q,O\}=O, \qquad \{Q,\tilde O\} \neq 0.
\eeq
This condition means that we consider only gauge invariant operators
which are defined up to a gauge transformation.\cite{12}

Correlation functions of interest are of the form
\beq
\langle O_1(x_1)\cdots O_n(x_n)\rangle,
\eeq
where $O_i$ are from the $Q$-cohomology. They are independent of
world-\-sheet and target space metric in the topological phase
\cite{2}. First of these properties means, in particular, that (2.6)
is independent on $x_1,\ldots,x_n$ and the second ensure its
independence of the coupling constant $1/r^2$. We will see later that
both of these properties are broken in the $\sigma$ model with the
black hole metric.

The $Q$-cohomology of observables in $d=2$ $\sigma$ model is
particularly simple. It consists of only two elements, one is the
partition function and another one can be chosen in the form
\cite{2,3}
\beq
O = iJ_{ij}(w)\chi^i\chi^j(x).
\eeq

Let us now consider correlation function (2.6) with operator $O$ from
(2.7). As it cannot depend on $r^2$ we can take limit $r^2
 \Rightarrow \infty$. Thus, the semiclassical approach becomes
exact. In particular,
 nonzero contributions come only from instantons (I) \cite{2}. The
latter are solutions of classical equations of motion in a given
topological class with winding number $k$. They are holomorphic
functions
\beq
w(z)=v\left(1+\sum^k_{\ell=1}\frac{\tilde
\rho_\ell}{z-z_\ell}\right),
\eeq
which satisfy the equation \beq
\prt_\mu w^i-\epsilon^\nu_\mu J^i_j\prt_\nu w^j =0,
\eeq
or $\bar \prt w=0$. Here $w=w^1+iw^2,$ $\bar w=w^1-iw^2.$
Instanton solution in (2.8) depends on $2k+1$ complex parameters:
$z_\ell$ are centers of multi-\-instanton, $\tilde \rho_\ell$
characterize its sizes and orientations, while $v$ is the overall
boundary condition at infinity. Hence, $I$ in (2.8) has $2k+1$ boson
zero modes \cite{2} $\prt w^i/\prt v $,
 $\prt w^i/\prt z_\ell $ and $\prt w^i/\prt\tilde \rho_\ell$,
 which correspond to variations with respect to
these parameters. However the one  associated with boundary
condition $v$ has  a quadratically divergent  norm on the world sheet
taken to be a complex plane. In fact parameter $v$ has a meaning of
vacuum expectation value (VEV) for field $w$. We are not going to
include the integration over $v$ in the instanton measure \cite{8}.
The reason is that we usually do not integrate over VEV in QFT. The
latter would mean summing up all the different vacuums of the theory.
Instead, we minimize the vacuum energy with respect to VEV to find
the true vacuum of the theory. Of course, if physics do not depend on
$v$ (like in topological $\sigma$ models with the compact target
space) than one could safety integrate over it \cite{2}; it makes
essentially no difference. However in the case of the $\sigma$ model
with black hole metric we are going to study here, the BRST symmetry
is broken and physics depends on $v$ as we will see later. Therefore,
we keep $v$ fixed.

Thus we are left with $2k$ boson zero modes to be included in the
instanton measure. Fermion zero modes are given by the same
expressions $\prt w^i/\prt z_\ell,$ $\prt w^i/\prt\tilde \rho_\ell$
as boson ones \cite{2,8}, since they are solution of the equation
\beq
\bar D\chi=0,
\eeq
which is identical to the equation for the boson zero modes. Hence,
we have $2k$ complex fermion zero  modes. This means that the
 correlation function
\beq
\langle O(x_1)\cdots O(x_n)\rangle,
\eeq
(here  $O$ is from eq.(2.7))
is nonzero  in the instanton background only if $n=2k$.
 In order to calculate it we
have to substitute (2.7) into (2.11), use expressions for fermion
zero modes $\chi^i$ and integrate over $z_\ell$, $\tilde \rho_\ell$
and over their fermion superpartners. The result can be written in an
elegant form. Instead of integration over $z_\ell, \tilde \rho_\ell$
let us proceed to new variables defined as follows. Fix points
$x_1\ldots x_n$ and consider $2k$ functions $w(x_1)\ldots w(x_n)$
given by (2.8) as functions of $z_\ell,\tilde \rho_\ell$. Then it is
easy to see that fermion zero modes $\chi^i$ (which are given by
$\prt w^i/\prt z_\ell$, $\prt w^i/\prt\tilde \rho_\ell$) represent
the jacobian needed to pass from variables $z_\ell,\rho_\ell$ to
$w(x_n)$. We get finally
\beq
\langle O(x_1)\ldots O(x_n)\rangle= g^k_I\int g(w_1)d^2 w_1 \ldots
g(w_n)d^2 w_n.
\eeq
Here $w_p=w(x_p),\, p=1\ldots n,$ while factors $g(w_p)$ arise from
factors $J_{ij}(w)$ in (2.7) when (2.2) is taken into account. For
the more detailed derivation of eq.(2.12) see \cite{8}.
The constant $g_I$ in eq.(2.12) is
\beq
g_I =e^{-S_I},
\eeq
where $S_I$ is the instanton action. In $\sigma$ models with compact
target space $S_I=0$, because the topological term in the instanton
action (the second term in r.h.s.
 of (2.1)) exactly cancels the kinetic
term (the first term in r.h.s.
 of (2.1)) for  holomorphic function $w$.
Thus $g_I=0$
and the independence of the correlation function (2.11) on
coordinates $x_1\ldots x_n$ as well as on the
coupling constant $1/r^2$ is
manifest in (2.12), provided the integrals are convergent.

Let us now consider the case of the black hole metric of the target
space. It has the form
\beq
g(w) = \frac 1{1+|w|^2}.
\eeq
Its  difference from, say, metric of the sphere for $O(3)$ $\sigma$
model
\beq
g^{sphere}(w) =\frac 1{(1+|w|^2)^2}
\eeq
in  its slow fall-off at large $|w|$. To see the relation of the
$\sigma$ model with metric (2.14) to the black hole let us perform
the
 change of variables
\begin{eqnarray}
w& = & \sinh r\, e^{-i\theta} \nonumber \\
\bar w &=& \sinh r\, e^{i\theta}.
\end{eqnarray}
The kinetic term in eq.(2.1) with the metric (2.14) becomes
\beq
S_{kin} = \frac{r^2}{2\pi}\int d^2x\left\{ (\prt_\mu r)^2
+\tanh r^2 (\prt_\mu \theta)^2\right\}.
\eeq
The latter is the familiar metric of the Euclidean black hole studied
in \cite{9} in the framework of gauged $SL(2R)/U(1)$ WZW model. Note
that gauged WZW model of ref. \cite{9} includes also the dilation term
which makes it conformal.

Let us now address a question \cite{13,8}: do holomorphic instantons
(2.8) still exist in the $\sigma$ model with metric (2.14). The main
problem is that the topological term becomes  logarithmically
divergent at large $w$. However the coefficient in front of the
logarithm does not depend on the regularization scheme and is still
proportional to the winding number $k$. To see this let us substitute
(2.8) into the second term in r.h.s. of (2.1). Using the metric from
(2.14) we have
\beq
S_{top}=\frac{r^2}\pi \int  d^2x\frac{(\bar \prt w\prt\bar w
-\bar \prt \bar w \prt w)}{1+|w|^2} =-2r^2k
\left[\log\frac 1a +\mbox{ const}\right],
\eeq
where we introduced the UV cutoff on the world sheet $1/a$ (lattice
spacing). Thus we still have the configuration space divided into
topological classes and holomorphic instantons (2.8) are still
minimum points of the kinetic term in a topological class with a
given winding number $k$. The explicit check that instanton (2.8) is
a solution to equation of motion is performed in ref.\cite{8}.

What about instanton action $S_I$ ? One may worry that instantons are
suppressed in the path integral if they have infinite action. It is
shown in \cite{8} that instanton (2.8) has nonzero but finite action
due to the cancellation between kinetic and topological terms in the
topological version of the $\sigma$ model (cf. ref.\cite{13} where
instantons are studied in the non-\-topological version of the
$\sigma$ model with the black hole metric and shown to have infinite
action). The result for $S_I$ is
\beq
S_I =k \frac{r^2}3.
\eeq
Hence, the constant $g_I$ in (2.12) becomes nontrivial
\beq
g_I= e^{-r^2/3}.
\eeq
It involves the dependence on $r^2$ which is the first signal for
BRST symmetry breaking.

Let us note that the black hole metric is the limiting case to have
unsuppressed instantons \cite{8}. If the divergence of the
topological term (2.18) were power rather than logarithm, then $S_I$
would be infinite \cite{8}.

Now let us consider the correlation function (2.11) in the $I$
background for the simplest case of $I$ with winding number $k=1$
\beq
w = v\left(1+\frac \rho{z-z_0}\right).
\eeq
(2.11) is nonzero only for two point correlation function $(n=2k=2)$.
Eq.(2.12) gives
\beq
\langle O(x_1)O(x_2)\rangle =g_I\int \frac{d^2w_1}{1+|w_1|^2|}
\frac{d^2 w_2}{1+|w_2|^2}.
\eeq
The integrals over modular space of $I$  become logarithmically
divergent in (2.22). Introducing UV and IR cutoff on the world sheet
$(1/a$ and $1/L$) we get with the double logarithmic accuracy
\cite{8}
\beq
\langle O(x_1)O(x_2)\rangle=2(2\pi)^2g_I\log\frac{|x_{12}|}a
\log\frac L{|x_{12}|}+O(\log),
\eeq
where $x_{12}=x_1-x_2$. The UV logarithm arises here when the
 instanton centre is close to either point $x_1$ or $x_2$ (either
$w_1$ or $w_2$ becomes large). The IR logarithm then comes from
the integration over $w_2$ or $w_1$ respectively and can be
rewritten as the logarithmic integral over the instanton size.

 The $x_{12}$-dependence in (2.23) means BRST
symmetry breaking. To see this explicitly consider $\prt_\mu O(x)$.
We have
\beq
\prt_\mu O=\left\{ Q,iJ_{ij}\prt_\mu w^i\chi^j\right\},
\eeq
 which means that $\prt_\mu O$ is $Q$-exact. Hence, nonzero value for
$\langle\prt_\mu O(x_1),O(x_2)\rangle$ means that
\beq
Q|0> \neq 0.
\eeq

Another  way to see BRST symmetry breaking is to calculate the vacuum
energy and to show that it is nonzero. Instantons by themselves
cannot produce nonzero $E_{vac}$ because of the anomalous selection
rule $n=2k$ ($I'$s have fermion zero modes). In ref.\cite{8}
instanton--\-anti-\-instanton pair $(I\bar I)$ was considered and
shown to produce nonzero $E_{vac}$.

$\bar I$ is an anti-holomorphic map from the world sheet to the
target
\beq
w = v\left(1 +\sum^p_{\ell=1} \frac{\bar \rho_{a\ell}}{\bar z-\bar
z_{a\ell}}\right),
\eeq
where $p$ is the winding number, $x_{a\ell}, \bar \rho_{a\ell}$ are
new complex parameters. In  nontopological versions of $\sigma$
models $I'$s and $\bar I'$s come on the same ground. Instead, in
topological $\sigma$ models with compact target space $I'$s come with
zero action, while
\beq
S_{\bar I} =2r^2pA,
\eeq
where $A$ is the area of the target space. The reason for
the result in (2.27) is
that the topological  term doubles the kinetic one for the
anti-\-holomorphic map.

On general grounds the dependence on $r^2$ cannot appear in
correlation functions if BRST symmetry is not broken. Thus, for
topological $\sigma$ models with compact target space $\bar I'$s
plays no role \cite{2}. Instead, for the topological $\sigma$ model
with black hole metric (2.14) $\bar I'$s do produce nonzero effects
\cite{8}.

Consider the $\bar I$ with winding number $-1$. Its action is
logarithmically divergent:
\beq
S_{\bar I} =4r^2\log \frac{|\rho_a|}a.
\eeq
Eq.(2.28) means that large size $\bar I'$s are suppressed in the path
integral. However small size $\bar I'$s (with size $|\rho_a|\sim a)$
induce a new point-\-like interaction. This is calculated in \cite{8}
in terms of an effective Lagrangian. The following vertex should be
added to the action (2.1)
\beq
V_{\bar I}=-g_{\bar I}\int d^2x g(w)\bar \chi\chi\,\prt^2_\mu
(g(w)\bar \chi\chi)
\eeq
in order to mimic the effect of $\bar I'$s. Here $g_{\bi}$ to be
treated together with $g_I$ as two new coupling constants of the
model. Four fermion fields in  (2.29) account for
four $\lambda^{\mu i}$
zero modes of $\bi$. From (2.29) it is clear that $V_{\bi}$ is
$Q$-exact and does not contribute to correlation functions if BRST
symmetry is not broken.

However for the case of the black hole metric the
vertex in(2.29) produces nonzero
$I\bi$ contribution to $E_{vac}$. Observe first that the calculation
of $E^{I\bi}_{vac}$ is essentially the same as the one for the
correlation function (2.22) in the one $I$ background. Eq.(2.29)
gives
\beq
E^{I\bi}_{vac}=\langle V_{\bi}\rangle_I=\left. -g_{\bi} \int
d^2_{x_1} \prt^2_{x_2}\langle O(x_1)O(x_2)\rangle
\right|_{x_2\to x_1}.
\eeq
Since correlation function $\langle O(x_1)O(x_2)\rangle$ shows
$x_{12}$-\-dependence for the case of black hole metric the
r.h.s.  in
(2.30) is nonzero. Substituting (2.23) into (2.30) we get with
logarithmic accuracy
\beq
E^{I\bi}_{vac} =-16\pi^2 g_Ig_{\bi} \frac V{a^2}\left\{ \log
\frac La +O(1)\right\},
\eeq
where $V$ is the volume of the world sheet.

The nonzero result for $E_{vac}$ in (2.31) confirms our
 conclusion that
BRST symmetry is broken for the $\sigma$ model with black hole
metric. However both results for correlation function in (2.23) and
for $E_{vac}$ in (2.31) contain IR logarithmic divergences which
 come
from the integration over the instanton size $\rho$. Our aim in this
paper is to sum up all the IR divergent effects to get the IR finite
results both for correlation functions and for $E_{vac}$. This allows
us to interpret what
physical degrees of freedom are "liberated" in eq.(2.23).

\section{Coulomb gas description}
\setcounter{equation}{0}

At the end of the previous section we calculated vacuum energy
induced by $I\bi$ pair. We have seen that it is IR divergent. The
divergence comes from the integration over size of $I$. This means
that instanton in a given $I\bi$ pair becomes of infinitely large
size. Thus the single instanton  approximation becomes invalid
because  $I'$s start to overlap and interact. In this section we are
going to consider the gas of $I\bi$ pairs and show that it is
actually a Coulomb gas. To do so we use the method of
instanton-\-induced effective Lagrangian \cite{14,8} which can be
applied to any  theory with instantons.

Let us first present the effective vertex for the single instanton
(2.21) with $k=1$. It has the form
\begin{eqnarray}
V_I&=&-\int d\mu_I \left(\frac{ir^2}2\right)^4 g^4(v)|v|^4\alpha_1
\lambda^{--}\bar \alpha_1\lambda^{++}\rho\alpha_2
\prt\lambda^{--}\bar \rho\bar \alpha_2 \bar \prt\lambda^{++}
\nonumber \\
&\times &\exp\left\{2r^2 g(v)[v\rho\prt\bar w
+\bar v\bar \rho\bar \prt w]\right\}.
\end{eqnarray}
Here $\alpha_1,\alpha_2$ are Grassmann variables which parametrize
fermion zero modes of $I$ (2.21)
\begin{eqnarray}
\chi_1& =&\frac{\alpha_1v}{z-z_0} \nonumber \\
\chi_2&=& \frac{\alpha_2 v\rho}{(z-z_0)^2},
\end{eqnarray}
while $d\mu_I$ is the instanton measure
\beq
d\mu_I =d^2x_0 d^2\rho d^2\alpha_1 d^2\alpha_2.
\eeq
The effective vertex (3.1) should be added to the action (2.1) to
mimic the effect of $I'$s at the perturbative level. To check it let
us calculate the following correlation function (cf.ref.\cite{8})
\beq
\langle w(x_1)\ldots w(x_n)\chi(x'_1)\chi(x'_2)\bar
\chi(x'_3)\bar \chi(x'_4)\rangle_I
\eeq
in the one $I$ background.

On one hand (3.4) can be calculated (in the leading order in $1/r^2)$
substituting classical expressions (2.21) and (3.2) for fields $w$
and $\chi$ into (3.4). This leads to
$$ \prod^n_{i=1}\left(v+\frac \rho{z_i-z_0}\right)
\frac{\alpha_1v}{(z'_1-z_0)}\frac{\alpha_2v\rho}{(z'_2-z_0)^2}
\frac{\bar \alpha_1\bar v}{(\bar z'_1-\bar z_0)}
\frac{\bar \alpha_2\bar v\bar \rho}{(\bar z'_2-\bar z_0)^2} $$
\beq
+\; \mbox{ permutations }\left(\begin{array}{l}
z'_1 \leftrightarrow z'_2 \\ \bar z'_1\leftrightarrow \bar z'_2
\end{array} \right).
\eeq
On the other hand, the same result can be reproduced in the purely
perturbative manner, inserting (3.1) into the action (2.1). Taking in
the expansion of $\exp-V_I$ the only first power in $V_I$ (this
corresponds to the one $I$ contribution) and taking into account
propagation functions
\begin{eqnarray}
\langle w(x), \bar w(0) \rangle &=& \frac 1{g(v)r^2}
\log\frac L{|x|}+v^2 \nonumber \\
\langle \bar \chi(x), \lambda^{++}(0)\rangle &=&
-\frac{2i}{g(v)r^2}\frac 1z, \nonumber \\
\langle \chi(x), \lambda^{--}(v)\rangle &=& -
\frac{2i}{g(v)r^2}\frac 1z,
\end{eqnarray}
one gets the same answer for correlation function (3.4) as in (3.5).

As the effective Lagrangian should depend on field $w$ rather on its
VEV we generalize (3.1) making the substitution $v\to w$ in (3.1).
This takes into account higher loop corrections to (3.1) (note that
we
actually derived $V_I$
 above in the one loop approximation). Making also
obvious generalization $\prt\lambda^{--}\to D\lambda^{--}$,
$\bar \prt\lambda^{++}\to \bar D\lambda^{++}$ and integrating over
Grassmann variables $\alpha_1,\alpha_2$ in (3.1) we get finally
\begin{eqnarray}
V_I&=&-\frac{g_Ir^8}{16}\int d^2xd^2\rho|\rho|^2 |w|^4 g^2(w)\,
g(w)\lambda^{--}\lambda^{++}g(w)D\lambda^{--}\bar D\lambda^{++}
\nonumber \\
&\times& \exp\{2r^2g(w)[\rho w\prt\bar w+
\bar \rho \bar w \bar \prt w]\}.
\end{eqnarray}
Now we have two effective vertices for $I$ and for $\bar I$ in eqs.
(3.7) and (2.29) respectively. These vertices determine, in
principle, the instanton physics in the model.

It is clear that nonzero contributions to vacuum energy can come only
from topologically trivial configurations with equal number $I'$s and
$\bi'$s. To study the medium containing both $I'$s and $\bi'$s we
will use $I\bi$ molecular gas approximation in this Section.
Expanding $\exp-(V_I+V_{\bi})$ in powers of $V_I$ and $V_{\bi}$ we
are going to contract fermion fields only in pairs. This gives
the $I\bi $ effective vertex:
\beq
V_{I\bi}(w)=-\langle V_I V_{\bi}\rangle.
\eeq
To give an idea what effects are taken into account in this
approximation and what are ignored consider, say, two $I$ -- two
$\bi$ contribution.

It is clear from (3.8) that graphs in which
  fermion lines connect $I$'s and $\bi$'s only inside pairs
 are taken into account, whereas those graphs in which each
$I$ and $\bi$ is
 connected  to other three by fermion lines are
ignored . This $I\bi$ approximation is valid provided the
instanton density is small, thus we assume $g_I\ll 1$, $g_{\bi}
\ll 1$ in this section.\footnote{This condition could be
insufficient to ensure the validity of $I\bi$ approximation if graphs
with connected fermion lines contain too strong IR divergences. We
will show in the next section that these graphs are IR finite.}

Let us now calculate the $I\bi$ effective Lagrangian (3.8). This
calculation is essentially of the same type as the one for $E_{vac}$
we have already performed in the previous section. Let us expand $w$
as $w=w_{ext}+w_{qu}$ in eqs. (3.7) and (2.29), where $w_{ext}$ is
the external field and $w_{qu}$ is the quantum fluctuation. Averaging
in (3.8) over fermions as well as over $w_{qu}$ using propagation
functions (3.6) we arrive at
\beq
V_{I\bi}=-8\pi g_Ig_{\bi}\int\frac{d^2x}{a^2}
\frac{d^2\rho}{|\rho|^2} e^{2r^2g(w)[\rho w\prt \bar w
+\bar \rho\bar w\bar \prt w]},
\eeq
where we put $w_{ext}\to w$ in the final equation. At $w=0$ (3.9)
reproduces our result for $E_{vac}$ (2.31). It is particularly clear
from (3.9) that the logarithmic divergence in (2.31) comes from the
integration over the instanton size $\rho$. Note that the size of
$\bi$ is small $|\rho_a|\sim a$. Moreover, the integral over $I\bi$
separation is dominated at small $|z_0-z_a|\sim a$, thus the typical
$I\bi$ configuration corresponds to the
very small $\bi$ located closely to
the center of large $I$. As we will see below the appearance of the
factor $d^2\rho/|\rho|^2$ is a signal for the Coulomb nature of
interactions in the gas of $I\bi$ pairs.

Let us compare (3.9) with the instanton  induced effective vertex of
$O(3)$ $\sigma$ model (the nontopological $\sigma$ model without
fermions with target space  metric (2.15)). It reads (cf. for example
\cite{13})
\beq
V^{O(3)}_I=-\mbox{ const }e^{-r^2}\int
\frac{d^2xd^2\rho}{|\rho|^4}\left(\frac{|\rho|}a\right)^b
e^{2r^2g^{sphere}(w)[\rho w\prt \bar w+\bar \rho \bar w
\bar \prt w]}.
\eeq
The first coefficient of $\beta$-function $b=2$ for $O(3)$ $\sigma$
model. Thus we have the same IR-\-logarithmic behaviour in (3.10) as
in (3.9). In $O(3)$ $\sigma$ model each $I$ can be represented as a
dipole of some "charge" and "anticharge" \cite{10} (instanton
quarks). These charges form the Coulomb gas system at the inverse
temperature $\beta=1$ \cite{10}, where $\beta$ is the coefficient in
the charge--\-anti-\-charge $(q\bar q)$ interaction potential
\beq
e^{-U_{+-}(x_1-x_2)} = e^{-2\beta\ln\frac{|x_1-x_2|}a}.
\eeq
The size of $I$ plays the role of the separation between charge and
anticharge. Thus, the factor $d^2\rho/|\rho|^2$ in (3.10) exactly
corresponds to $\beta=1$ in (3.11).

This temperature is above the point of Kosterlitz-\-Thouless phase
transition \cite{15} $(\beta=2)$, hence, the Coulomb gas is in the
plasma state \cite{16}. This means that the dynamically generated
mass scale appears due to the Debye screening mechanism and all the
IR divergences disappear. This "instanton induced" Coulomb gas is in
fact equivalent to the SG theory \cite{16}.

As the $I\bi$ vertex in (3.9) is of the same form as $I$ vertex in
(3.10), we conclude that the $I\bi$ pair in the black hole model plays
the same role as a single $I$ in $O(3)$ $\sigma$ model. Hence, the
gas of $I\bi$ pairs in the black hole model represents the Coulomb
plasma at the inverse  temperature $\beta=1$.

This can be verified directly without reference to $O(3)$ $\sigma$
model using $II$ vertex (3.10). For example, the interaction
potential for two $I\bi$ pairs
\beq
U^{(I\bi)^2}=\langle 2r^2g(w)[\rho_1w\prt\bar w+\bar
\rho_1\bar w\bar \prt w](x_1), 2r^2g(w)[\rho_2 w\prt
\bar w +\bar \rho_2\bar w\bar \prt w](x_2)\rangle
\eeq
can be compared with that for the Coulomb system of two charges and
two anticharges at the locations $x_1,$ $x_1+\rho_1,x_2,$
$x_2+\rho_2$. Classically (in the  leading order in $r^2$ )
(3.12) is zero. Next-\-to-\-leading corrections in (3.12) can be
analyzed and shown to reproduce the desirable Coulomb potential
independently of $g(w)$, provided $g(w)\to0$ if $|w|\to\infty$. We
are not going to do it here.

The arguments above lead us to the conclusion that $I\bi$ gas in the
black hole $\sigma$ model in $I\bi$ approximation can be described by
the SG effective action
\beq
S^{(b)}_{eff}=\frac 1\pi\int d^2x\left\{\frac 12(\prt_\mu\phi)^2-
\frac{2\pi g_q}{a^2}\cos 2\phi\right\},
\eeq
where $\phi$ is a real scalar field, $g_q$ is a "fugacity" of charges
to be determined below. To fix $g_q$ in terms of $g_Ig_{\bi}$ from
(3.9), let us calculate the $q\bar q$ contribution to the vacuum
energy and compare the result with $E_{vac}^{I\bi}$ in (2.31). We
have
\begin{eqnarray}
E^{q\bar q}_{vac}&=&-g^2_q\int \frac{d^2x_1}{a^2}\frac{d^2x_2}{a^2}
\langle e^{2i\phi(x_1)}, e^{-2i\phi(x_2)}\rangle_{\mbox{ free boson}}
\nonumber \\
&=&-g^2_q\frac V{a^2}\int \frac{d^2\rho}{|\rho|^2} =-2\pi\,
g^2_q\frac V{a^2}\log\frac La,
\end{eqnarray}
where $\rho=x_1-x_2$.

Note that we have got an IR logarithm in (3.14) because we have
considered the contribution of a single $q$ --- single $\bar q$ to
$E_{vac}$. The vacuum energy of the CG/SG system
(3.13) is IR-\-finite, because field $\phi$ acquires a dynamically
generated mass in (3.13), as we note above. We postpone the
calculation of $E_{vac}$ for the black hole model till section 5.

Comparing $E_{vac}^{q\bar q}$ with $I\bar I$ vacuum energy (2.31) we
get
\beq
g^2_q=8\pi g_I g_{\bi}.
\eeq

Now let us relate field $w$ to field $\phi$ from the effective action
(3.13). Using again the equivalence of $I$ gas in $O(3)$ $\sigma$
model and $I\bi$ gas in the topological black hole model, we can
learn from \cite{10} that
\begin{eqnarray}
w & = & e^{-i(\phi-\phi*)} \nonumber \\
\bar w &=& e^{-i(\phi+\phi*)},
\end{eqnarray}
where $\phi*$ is the dual field to $\phi$:
\beq
\prt_\mu \phi* =i\epsilon_{\mu\nu}\prt_\nu \phi.
\eeq
To make sense of (3.16) we assume that the constant mode of
$\phi$ is analytically continued to the imaginary values.

Let us check eq.(3.16). To do so consider field $w$ produced by the
single $q\bar q$ and compare the result with that for the field $w$
of a single $I\bar I$ pair. Eq. (3.16) gives
\beq
\langle w(x)\rangle_{q\bar q}=\langle e^{-i(\phi-\phi*)(x)},
e^{2i\phi(x_0)} e^{-2i\phi(x_0-\rho)}\rangle_{\mbox{ free boson} },
\eeq
where we represented charge and anti-charge by exponentials using
(3.13). It is easy to check with the help of the definition of the
dual field (3.17) that the propagation function of a chiral part
$(\phi-\phi*$) of field $\phi$ is given by
\beq
\langle(\phi-\phi*)(x),\phi(0)\rangle=\frac 12\ln \frac Lz.
\eeq
Substituting (3.19) into (3.18) we arrive at
\beq
\langle w\rangle_{q\bar q}=\left(1+\frac \rho{z-z_0}\right)
e^{-i(\langle\phi-\phi*\rangle)},
\eeq
where $\langle\phi-\phi*\rangle$ is the classical VEV of the field
$\phi-\phi*$.

What about the field of a single $I\bar I$ pair ? From (3.9) it is
clear that $\langle w\rangle_{I\bi}$ coincides with instanton
solution (2.21) in the leading order in $r^2$. Comparing (3.20) and
(2.21) we see that they do coincide with the natural identification
\beq
v=e^{-i\langle\phi-\phi*\rangle},
\eeq
which relates the classical VEV's of fields $w$ and $\phi$ in
accordance with (3.16).

To sum up, the above results mean that in the leading approximation
at $g_I\ll 1$, $g_{\bi}\ll 1$, and $r^2\gg 1$ the topological
$\sigma$ model with black hole metric is equivalent to the sine
Gordon theory (3.13) with $g_q$ given by (3.15). Any correlation
function of field $w$ can be expressed (in the same approximation) in
terms of correlation functions of SG model. Namely,
\beq
\langle F_1[w(x_1)],\ldots F_n[w(x_n)]\rangle_{BH}=\frac{
 \langle
F_1[e^{-i(\phi-\phi*)(x_1)}],\ldots
F_n[e^{-i(\phi-\phi*)(x_n)}]\rangle_{SG}}
{ \langle
F_1[e^{-i(\phi-\phi*)(x_1)}],\ldots
F_n[e^{-i(\phi-\phi*)(x_n)}]\rangle_{free\; boson}},
\eeq
where $\langle\ldots\rangle_{BH}$ and $\langle\ldots\rangle_{SG}$
mean correlation functions in black hole and SG models
respectively. However in the
topological $\sigma$ model we are interested
in correlation functions of type (2.11).
 In section 6 we will express these
 in terms of
correlation functions of the SG model.

To conclude this section let us rewrite the effective theory (3.13)
in terms of fermions to make it more practical in calculations. At
the point $\beta=1$ SG model (3.13) is equivalent to free
massive fermions \cite{16,17}
\beq
S^{(f)}_{eff} = \frac 1\pi \int d^2x\{ \bar \psi i\gamma_\mu
\prt_\mu \psi + im\bar \psi\psi\}.
\eeq
Here $\psi$ is a two component spinor $\psi=\left({\psi_1 \atop
\psi_2}\right)$, while $\gamma_1=\sigma_1,$ $\gamma_2=\sigma_2$
($\sigma_1$ and $\sigma_2$ being Pauli matrices). The relation of the
fermion field $\psi$ to  SG field reads \cite{18}
\beq
\begin{array}{ll}
\psi_1=c:e^{-i(\phi-\phi*)}:, & \bar \psi_2=c:e^{i(\phi-\phi*)}:,\\
\bar \psi_1=c:e^{-i(\phi+\phi*)}:, & \psi_2=c:e^{i(\phi+\phi*)}:
\end{array}
\eeq
The normalization constant $c$ can be fixed comparing, say,
propagation function $\langle\psi_1,\bar \psi_2\rangle$ in models
(3.13) and (3.23):
\beq
c^2= \frac 1{2ia}.
\eeq
Rewriting the mass term in (3.23) in components
\beq
\frac 1\pi \int d^2x\{im \bar \psi_1\psi_1+im \bar \psi_2\psi_2\}
\eeq
and comparing it with (3.13) we can fix the value of the fermion mass
in terms of bare coupling constant $g_q$
\beq
m=-\frac{2\pi g_q}a.
\eeq
Thus our black hole model is equivalent to free massive fermions
(3.23) at $g_I\ll 1$, $g_{\bi}\ll 1$ and $r^2\gg 1$. In the next
section we will relax first two of these conditions, showing that
$I\bar I$ approximation becomes exact provided a proper definition of
the geometry of the instanton modular space is used. As for
perturbative corrections in $1/r^2$ to the partition function and to
correlation functions of observables (2.11), they should be zero
since the theory is topological at the perturbative level. In other
words, the dependence of observables on $r^2$ could come only in the
combinations $g_I$ and $g_{\bi}$, because BRST symmetry is broken
only by instanton effects. However we have no rigorous proof of this
assertion here.

\section{Exact $(I\bi)^2$ calculation}
\setcounter{equation}{0}

In the previous section we showed that the black hole model has
CG/SG description in the
$I\bi$ approximation. Here we will calculate
 the contribution of a two $I\bi$ pairs to the partition
function exactly and compare the result with that given by
the $I\bi$-approximation
. Our aim is to study the possible corrections to the
Coulomb gas picture.

Like $I\bi$ contribution to the partition function (2.30) the
$(I\bi)^2$ one can also be expressed in terms of correlation function
(2.11) in the instanton background. Eq. (2.29) for $V_{\bi}$ gives
\beq
Z^{(I\bi I\bi)}=\frac 12 g_{\bi}\int d^2x_1 d^2x_3
\prt^2_{x_2}\prt^2_{x_4}\langle O(x_1)O(x_2)O(x_3)O(x_4)
\rangle_{x_2=x_1 \atop x_4=x_3}.
\eeq
According to our selection rule, $n=2k$, thus the
 correlation function
in r.h.s.
 of (4.1) is nonzero only for $I$ with winding number $k=2$
(see (2.8))
\beq
w(z)=v\left[1+\frac{\tilde \rho_1}{z-z_1}+ \frac{\tilde
\rho_2}{z-z_2} \right].
\eeq
For this correlation function the general eq.(2.12) gives
\beq
\langle O(x_1)O(x_2)O(x_3)O(x_4)\rangle = g^2_I \int
\frac{d^2 w_1}{1+|w_1|^2}\frac{d^2w_2}{1+|w_2|^2}
\frac{d^2w_3}{1+|w_3|^3}\frac{d^2w_4}{1+|w_4|^2},
\eeq
where $w_1\ldots w_4$ are values of (4.2) in the points $x_1\ldots
x_4$.

Let us first analyze (4.1) in the leading logarithmic approximation.
The calculation is quite similar to the one for $E^{I\bi}_{vac}$ in
(2.31). Two of the four logarithmic factors in (4.3) gives
$\delta(0)\sim1/a^2$ after operators  $\prt^2$ are applied in (4.1).
An other two can be written down
as logarithmic integrals over $\tilde
\rho_1, \tilde \rho_2$ (cf. eq.(3.9)). We have
\beq
Z^{(I\bi I\bi)}=\frac 12(2\pi)^2 g^2_I g^2_{\bi}
\frac{16}{a^4} V\int \frac{d^2x_{13}d^2\tilde \rho_1 d^2\tilde
\rho_2}{|\tilde \rho_1|^2 |\tilde \rho_2|^2},
\eeq
where $x_{13}=x_1-x_3$. Performing the $\tilde \rho$ integrals in
(4.4) we arrive with double logarithmic accuracy at
\beq
Z^{(I\bi I\bi)}=\frac 12 (16\pi)^2 g^2_I g^2_{\bi}
\frac{V^2}{a^4}\log^2 \frac La.
\eeq
The result in (4.5) should be expected. It is nothing other than the
second nontrivial term in the expansion of the partition function
$Z=\exp(-E_{vac})$ with $E_{vac}$ taken in the $I\bi$-approximation
(2.31).

The next question we are going to address concerning $Z^{(I\bi I\bi)}$
in (4.1), (4.3) is the presence of the corrections to (4.5) of the
type
\beq
g^2_Ig^2_{\bi} \frac V{a^2} \log \frac La.
\eeq
The appearance of  such a term would provide the $O(g^2_Ig^2_{\bi})$
correction to $E^{I\bi}_{vac}$ in (2.31).
This would mean the presence of an extra IR divergence, besides the
ones taken into account in the $I\bi$ approximation.

To study the possible corrections of type (4.6) one can subtract the
double logarithmic term (4.5) from (4.1) and look for the
logarithmic divergence in the rest integral. The answer is that there
are no dangerous contributions of type (4.6) in $Z^{I\bi I\bi}$. This
means that there are no new IR divergences in our model. In other
words, all the IR divergences are taken into account in the $I\bi$
approximation.

Now let us turn to non-logarithmic contributions to (4.1). To
calculate a constant correction to logarithm one has to write down a
more accurate expression for $Z^{I\bi I\bi}$ instead of (4.4).
However this is not very useful because the region of integration in
(4.4) is not defined with appropriate accuracy. To give an example of
such uncertainty let us relate parameters $\tilde \rho_1, \tilde
\rho_2$ in (4.4) with instanton sizes $\rho_1,\rho_2$ which appear in
the effective action approach of the previous section.

Using effective vertex (3.9) we can calculate the field of two
$I\bi$ pairs as $\langle w(x),V_{I\bi}V_{I\bi}\rangle$. We have
\beq
\langle w(x)\rangle_{I\bi I\bi}=\left(1+
\frac{\rho_1}{z-z_1}\right)\left(1+\frac{\rho_2}{z-z_2}\right).
\eeq
Comparing this with (4.2) (note, that instanton field coincides with
that of $I\bi$ pair at $r^2\gg1$) we have
\begin{eqnarray}
\tilde \rho_1&=&\rho_1\left(1+\frac{\rho_2}z\right), \nonumber \\
\tilde \rho_2&=&\rho_2\left(1-\frac{\rho_1}z\right),
\end{eqnarray}
where $z=z_1 -z_2$
 is the distance between instanton centres. Observe now that
constant corrections to (4.4) depend on the definition of the region
of integration in (4.4). Say, we can choose the region
\begin{eqnarray}
a & \leq & | \tilde \rho_1| \leq L, \nonumber \\
a & \leq & | \tilde \rho_2 |\leq L,
\end{eqnarray}
or instead
\begin{eqnarray}
a & \leq & |\rho_1| \leq L, \nonumber \\
a & \leq & | \rho_2 | \leq L,
\end{eqnarray}
or make some other choice. In the $I\bi$ approximation of the last
section the choice (4.10) arises naturally, since $(I\bi)^2$ effect
appears as a $V^2_{II}$  term in the expansion of $\exp(-V_{I\bi})$,
thus $\rho_1$ and $\rho_2$ come as an independent parameters.

We see that our model needs a more accurate definition of the
geometry of the modular space of instantons. One can think of
eq.(4.4) as being exact, and define the region of integration in it.
We choose the one in (4.10).

The motivation is as follows. Suppose, instead of (4.10) we take
(4.9) as the region of integration in (4.4). Then the answer (4.5)
for $Z^{I\bi I\bi}$ becomes exact. This means that the contribution
$O(g^2_Ig^2_{\bi}VL^2/a^4)$ which is needed to solve the IR problem is
absent in $Z^{I\bi I\bi}$. The common belief is that the field theory
with IR divergences is not reasonable. One has to use the freedom in
the definition of the geometry of modular space of instantons to get
a IR-\-finite theory. We will show below that the choice (4.10) will
solve the IR problem.

To make (4.4) more transparent let us proceed from the integration
over $\tilde \rho_1$, $\tilde \rho_2$ to the one over $\rho_1,\rho_2$
and assume that the region of integration is (4.10). Using (4.8) we
get
\beq
Z^{I\bi I\bi}=\frac 14(2\pi)^2g^2_I g^2_{\bi} \frac{16}{a^4} V
\int\frac{d^2x_{13}d^2\rho_1d^2\rho_2|x_{13}|^2
|x_{13}+\rho_2-\rho_1|^2}{|\rho_1|^2|\rho_2|^2|x_{13}+\rho_2|^2
|x_{13}-\rho_1|^2},
\eeq
where we replace $z$ by $x_{13}$ because one $I$ comes close to $x_1$
and another to $x_3$. From (4.11) it is clear that we have to add two
extra conditions to (4.10) in order to cut the integral (4.11) in the
ultra-\-violet region. Namely, we have to impose
\begin{eqnarray}
a & \le& |x_{13} +\rho_2| \le L, \nonumber \\
a &\le & |x_{13}-\rho_1|\le L.
\end{eqnarray}
We will see below that the symmetry $\rho_1\leftrightarrow x_{13}
+\rho_2$; $\rho_2\leftrightarrow x_{13}-\rho_1$ reflects the
possibility of interchanging
 positions for anti-\-charges in the $(q\bar
q)^2$ system in the Coulomb gas description.

Let us compare (4.11) with the result for $Z^{I\bi I\bi}$ in $I\bi$
approximation. In this approximation $(I\bi)^2$ contribution to $Z$
is equal to the contribution of two $q\bar q$ pairs to the partition
function. Using (3.13) we have
\begin{eqnarray}
Z^{q\bar q q\bar q}&=& \frac 14 g^4_q\int \frac{d^2x_1}{a^2}
\frac{d^2\rho_1}{a^2}\frac{d^2x_3}{a^2}\frac{d^2\rho_2}{a^2} \\
&\times& \left\langle e^{2i\phi(x_1)}e^{-2i\phi(x_1-\rho_1)}
e^{2i\phi(x_3)} e^{-2i\phi(x_3-\rho_2)}
\right\rangle_{\mbox{ free boson }}, \nonumber
\end{eqnarray}
where charges and anti-charges are
taken to have coordinates $x_1,x_3$
and $x_1-\rho_1$, $x_3-\rho_2$ respectively. Using the relation
\beq
\left\langle e^{2i\phi(x)}e^{\pm2i\phi(y)}
\right\rangle_{\mbox{ free boson }}
=\left(\frac{|x-y|}a\right)^{\pm2}.
\eeq
It is easy to see that (4.13) gives the same integral as in (4.11).
We conclude therefore that with the definition (4.10), (4.12) the
$I\bi$ approximation becomes exact for $Z^{I\bi I\bi}$. Furthermore,
it is not difficult to check that the definition of the geometry of
the modular space of instanton for any winding number $k$ in terms of
sizes $\rho_1,\ldots\rho_k$ like the one in (4.10), (4.12)  for $k=2$
makes our topological black hole model equivalent to the CG/SG system
(3.13) exactly at any values of $g_I, g_{\bi}$. Of course, this makes
it IR-\-finite.

\section{Vacuum energy}
\setcounter{equation}{0}

The $I\bi$ vacuum energy (2.31) shows the IR divergence. In this
section we re-\-examine the calculation of the vacuum energy in the
black hole model using the sine Gordon/free fermions description
(3.13),(3.23).

First, let us minimize the classical vacuum energy in (3.13). The
vacuum state can be chosen at
\beq
\phi_0 =0.
\eeq
This means in accordance with (3.16) or (3.21) that
\beq
v = e^{i\alpha}
\eeq
with $\alpha$ real. The reason for this result is that
  the constant mode of $\phi*$ is not fixed, see
(3.17). Thus, as we mentioned before, the physics in the model is
$|v|$ dependent and the minimization of the vacuum energy gives
\footnote{ Note that this result is different from that in
ref.\cite{8}. It comes here
as a consequence of a CG/SG description of the
instanton vacuum.}
\beq
\langle |w|\rangle = |v|=1.
\eeq

What about the U(1) symmetry $w\to e^{i\gamma}w$ of the model.
Eq.(5.3) shows that U(1) symmetry is not broken. To check this we can
calculate the propagation function of "would-\-be" Goldstone boson.
If U(1) were broken the phase of field $w$ would be massless. Phase
of $w$ is related to $e^{2i\phi*}$, see (3.16).

Consider
\beq
\langle e^{2i\phi*(x)} e^{-2i\phi*(y)}\rangle =
(2ia)^2
\langle\psi_1\psi_2(x), \bar \psi_1\bar \psi_2(y)\rangle,
\eeq
where we use the correspondence (3.24). The propagation function of a
free massive fermion reads
\beq
\langle\psi(x)\bar \psi(0)\rangle=\frac{im}2\left\{\left(
\begin{array}{cc} 0 & \frac{\bar z}{|x|} \\ \frac{z}{|x|}
 & 0 \end{array}
\right) K'_0(m|x|)+\left(\begin{array}{ll} 1 & 0 \\ 0 & 1\end{array}
\right) K_0(m|x|)\right\},
\eeq
where $K_0$ is the zero order modified Bessel
 function and $K'_0$ is its
derivative with respect to the argument. Using (5.5) we get
at $ |x-y|\to\infty$
\beq
\langle e^{2i\phi(x)}e^{-2i\phi*(y)}\rangle
=(2\pi)^2g_q^2[K_0^{'2}(m|x-y|)-K_0^2(m|x-y|)]
\sim \frac{e^{-2m|x-y|)}}{(m|x-y|)^{3/2}}.
\eeq
Thus we conclude that there is no Goldstone boson in the model: the
phase of $w$ couples to the same massive fermion field as $|w|$ does.

Let us now consider the vacuum energy of our model in quantum theory.
It is known that the SG model is renormalizable at $\beta\le2$
 \cite{17}.
All UV divergences can be associated with the renormalization of the
coupling constant $g_q$, except vacuum energy. The
vacuum
energy is associated with
 another divergence at $1\le\beta\le2$. Let us calculate
it. We have in the fermion description
\beq
E_{vac}=\frac 12 \frac{im}\pi \int d^2x\langle\bar \psi_2\psi_2 +
\bar \psi_1\psi_1 \rangle.
\eeq
Plugging (5.5) into (5.7)  we arrive at
\beq
E_{vac}=-\frac{m^2}{2\pi}\, V\log \frac 2{ma}.
\eeq
Let us compare (5.8) with (3.14) or (2.31). The UV logarithm
$\log1/a$ comes in (5.8) as a single $q\bar q$ effect like in (3.14),
while the IR cutoff at $1/m$ arises due to the Debye screening in the
Coulomb plasma. Note, that $E_{vac}$ in (5.8) is still non-\-zero and
implies the BRST symmetry breakdown.

\section{Correlation functions}
\setcounter{equation}{0}

Now let us turn to the calculation of the correlation functions of
observables (2.11) which interest us in the TFT. In this section we
consider the two point correlation function (2.22) and show that it
is IR-\-finite. As our SG effective boson action (3.13) arises as a
result of integration over fermions $\chi$ and $\lambda$ we expect
that the correlation function of fermion fields (2.22)
could appear to be a complicated object
(non-\-local object, in fact) in terms of the field of the SG model.

The two point correlation function (2.22) is non-zero in the
background of a single $I$ plus an arbitrary number of $I\bar I$
pairs. Using the effective vertex (2.29) for $\bar I$ we get the
general expression
\beq
\langle O(x_1)O(x_2)\rangle_{I+(I\bar I)^n}= \frac{g^n_{\bi}}Z
\int \prod^n_{i=1}d^2 y_i\partial^2_{y_i'}\langle O(x_1)O(x_2)
\prod^n_{i=1}O(y_i)O(y'_i)\rangle_{y'_i=y_i}
\eeq
for (2.22) in the background $I+(I\bi)^n$.

To simplify life let us consider $I+I\bi$
contribution in (6.1) as an example,  like in section 4.
 Then (6.1) takes the form
\beq
\langle O(x_1)O(x_2)\rangle_{I+I\bi}=\frac{g_{\bi}}{Z}\int d^2y
\partial^2_{y'}\langle O(x_1)O(x_2)O(y)O(y')\rangle_{y'=y}.
\eeq

Here the four point correlation function is given by (4.3). The
calculation is quite similar to that for $Z^{I\bi I\bi}$
in section 4. One of the four logarithms in (4.3) gives $\delta(0)
\sim1/a^2$ under the action of $\partial^2_{y'}$ in (6.2), while
another is ultraviolet and gives $\log|x_{12}|/a$, like in (2.23).
The remaining two can be written down as  logarithmic integrations
over $\tilde \rho_1$ and $\tilde \rho_2$, like in (4.4). We get
\beq
\langle O(x_1)O(x_2)\rangle_{I+I\bi} =8(2\pi)^2 g^2_I
g_{\bi}\log\frac{|x_{12}|}a \int\frac{d^2yd^2\tilde \rho_1 d^2\tilde
\rho_2}{a^2(|\tilde \rho_1|^2+|x_{12}|^2)|\tilde \rho_2|^2}.
\eeq
Introducing new variable $\tilde \rho'_1$ with $|\tilde
\rho'_1|^2=|\tilde \rho_1|^2+|x_{12}|^2$ we can rewrite the integral
over $\tilde \rho_1$ in (6.3) as
\beq
\int \frac{d^2\tilde \rho'_1}{|\tilde \rho'_1|^2}
\eeq
over the region
\beq
|\tilde \rho'_1|\ge |x_{12}|.
\eeq

Let us now proceed from integration over $\tilde \rho_1', \tilde
\rho_2$ to the one over $\rho_1$, $\rho_2$ defined as in (4.8) with
$z=y-x_1$ (in fact $"1/2\log|x_{12}|/a"$ arises when the centre of
one of $I'$s close to $x_1$ and another $"1/2\log|x_{12}|/a"$ arises
when it closes to $x_2$. We take $z=y-x_1$ for simplicity: formulas
below should be understood as a symmetrization with respect to
$x_1\rightleftharpoons x_2)$. After changeing variables we get
\begin{eqnarray}
\langle O(x_1)O(x_2)\rangle_{I+I\bi}&=&
Z^{-1}8(2\pi)^2 g^2_I g_{\bi}\log\frac{|x_{12}|}a \\
&\times& \frac 12\int
\frac{d^2zd^2\rho_1d^2\rho_2|z|^2|z-\rho_1+\rho_2|^2}{|\rho_1|^2|
\rho_2|^2|z-\rho_1|^2|z+\rho_2|^2}. \nonumber
\end{eqnarray}
This is the representation for Coulomb system of $(q\bar q)^2$ with
charges at points $x_1,$ $x_1+z=y$ and anti-\-charges at points
$x_1-\rho_1,$ $y-\rho_2$ (see eqs. (4.11),(4.13)).
We have
\begin{eqnarray}
\langle O(x_1)O(x_2)\rangle_{I+I\bi}&=&
2(2\pi )g_I g_q g_{\bar{q}}log\frac{|x_{12}|}{a}\frac 12
\int \frac{d^2 y}{a^2} \frac{d^2 \rho_1}{a^2}
 \frac{d^2 \rho_2}{a^2}\\
&\times& \left\langle e^{2i\phi(x_1)}e^{-2i\phi(x_1-\rho_1)}
e^{2i\phi(y)} e^{-2i\phi(y-\rho_2)}
\right\rangle_{free\; boson}, \nonumber
\end{eqnarray}
where we made a distinction between $g_q$ and $g_{\bar{q}}$
for a moment. The result in (6.7) is nothing other but
the  $O(g_I g_q^2)$ term in the expansion of the
following correlation function of the SG model:
\begin{eqnarray}
\langle O(x_1)O(x_2)\rangle &=&
2(2\pi )\frac{g_I}{ g_{\bar{q}}}log\frac{|x_{12}|}{a}
\int^{g_q} d\bar{g_q}\int  \frac{d^2 \rho_1}{a^2} \\
&\times& \left\langle e^{2i\phi(x_1)}e^{-2i\phi(x_1-\rho_1)}
\right\rangle_{SG}. \nonumber
\end{eqnarray}
Eq. (6.8) is our desired expression of the correlation function
(2.22) in terms of a correlation function of the SG model.

What about the constrain (6.5)? Can it be rewritten down as
any condition on correlation functions of SG model?
Consider the correlation function in the r.h.s. of (6.8).
It can be written down in the form
\begin{eqnarray}
 \left\langle e^{2i\phi(x_1)}e^{-2i\phi(x_1-\rho_1)}
\right\rangle_{SG} & = &
\sum_{n=1}^{\infty} \frac{g^{2n}_q}{(n!)^2}\int
\prod_{i=1}^{n} \frac{d^2 y_i}{a^2} \frac{d^2 \rho_i}{a^2}
\frac{a^2}{|\tilde{\rho_{1+}}|^2}
\frac{a^2}{|\tilde{\rho_{1-}}|^2}
\frac{|\rho_1|^2}{a^2}\\
& \times & \frac{1}{Z_{SG}}
 \left\langle \prod_{i=1}^{n} e^{2i\phi(y_i)}e^{-2i\phi(y_i-\rho_i)}
\right\rangle_{free\; boson},\nonumber
\end{eqnarray}
where we introduced
\beq
\tilde{\rho_{1+}}=\rho_1\prod_{i=1}^{n}\frac{x_1-y_i+\rho_i}
{x_1-y_i},
\eeq
while
\beq
\tilde{\rho_{1-}}=\rho_1\prod_{i=1}^{n}\frac{x_1-y_i-\rho_1}
{x_1-y_i-\rho_1+\rho_i}
\eeq
Here (6.10) is the obvious generalization of the expression
for $\tilde{\rho'_{1}}$ in (6.5) (see (4.8)) for arbitrary $n$
, while (6.11) is the same quantity as in (6.10) with all
charges and anti-charges interchanged. Generalizing the
condition (6.5) to the arbitrary $n$, we get
\begin{eqnarray}
|\tilde{\rho_{1+}}| & \ge & |x_{12}|, \\
|\tilde{\rho_{1-}}| & \ge & |x_{12}| \nonumber
\end{eqnarray}
Substituting (6.13) into (6.9) we arrive finally at the constrain
\beq
 \left\langle e^{2i\phi(x_1)}e^{-2i\phi(x_1-\rho_1)}
\right\rangle_{SG} \le \frac{a^2|\rho_1|^2}{|x_{12}|^4}
\eeq
Eq. (6.13) gives us the lower bound of $\rho_1$ in the integral over
 $\rho_1$ in the r.h.s. of (6.8). Note, that we considered
the $I+I\bi$ system above only as a simplifing example.
All the steps of the calculation leading to eqs.(6.8), (6.13) can
be repeated for arbitrary $n$.

Let us now calculate the correlation function (6.8). Using the
fermion representation we get
\begin{eqnarray}
\langle e^{2i\phi(x)} e^{-2i\phi(y)}\rangle & = &
(2ia)^2
\langle\bar{\psi_2}\psi_2(x), \bar \psi_1 \psi_1(y)\rangle \\
& = & (2\pi)^2g_q^2K^{'2}_0(m|\rho_1|). \nonumber
\end{eqnarray}
Consider first the limit of  small distances $m|x_{12}|\ll 1$.
Approximating $K'_0$ in this limit as
\beq
K'_0(x)=-\frac 1x +O(x)
\eeq
and using eq.(3.27) we get from the constrain (6.13)
\beq
|\rho_1|\ge |x_{12}|.
\eeq
Performing the integral over $\rho_1$ in (6.8) in these limits
we finally arrive at
\beq
\langle O(x_1),O(x_2)\rangle=2(2\pi)^2g_I\log\frac{|x_{12}|}a
\log\frac 2{m|x_{12}|}.
\eeq
Comparing this with (2.23) we see that the only modification is the
appearance of IR cutoff parameter $1/m$. (6.17) still shows the
$x_{12}$-dependence of $\langle O(x_1)O(x_2)\rangle$ which is a
signal for BRST symmetry breaking.

Now let us estimate (6.9) at large distances $m|x_{12}|\gg 1.$
In this limit the correlation function in (6.14) shows the
exponential fall-off:
\beq
 \left\langle e^{2i\phi(x_1)}e^{-2i\phi(x_1-\rho_1)}
\right\rangle_{SG}\sim e^{-2m|\rho_1|}.
\eeq
We assumed that $m|\rho_1|\gg 1$ here. We will check it is true
below. Plugging (6.18) into the constrain (6.13) we get
\beq
|\rho_1|\ge \frac 1m log(m^2|x_{12}|^2).
\eeq
In particular eq.(6.19) shows that  $m|\rho_1|\gg 1$, indeed.
Integrating over $\rho_1$ in (6.8) within the limits determined
by (6.19) we get at  $m|x_{12}|\gg 1$
\beq
\langle O(x_1),O(x_2)\rangle \sim \frac 1{m^4|x_{12}|^4}
\eeq
We see that (6.20) shows a power fall-off
 at $|x_{12}|\rightarrow
\infty$. This means the presence of massless particles in our
theory.

As we showed above our topological black hole $\sigma$ model
is equivalent to free massive fermions. One may worry therefore,
how the power behaviour in (6.20) can appear.
Technically it comes because  correlation functions of the
fermion field $\chi$ of the original model is expressed in a non-local
manner in terms of correlation functions of SG model. Let us
note however, that from the physical point of view the
result in (6.20) could be expected.
We
interpret this behaviour as a propagation of the goldstino fermion
which appears as a consequence of spontaneous BRST-symmetry breaking.

To see this, observe that the condition (2.25) means the existence of
a (composite) goldstino fermion $\psi_g\sim Q|0>$.
It is easy to see making the $Q$-transformation in the initial action
of the model that this fermion couples to the current
\beq
\bar \psi_{g_\mu} \sim J_{ij}\partial_\mu w^i\chi^j.
\eeq
Observe now  that the r.h.s.
 of (6.15) is related to $\partial_\mu O(x)$
according to (2.24). Hence, we have
\beq
\partial_\mu O\sim \bar \psi_{g\mu}\psi_g.
\eeq
We see that the operator $O$ (2.7) couples to the goldstino mode.
Hence, the correlation function $\langle O(x_1)O(x_2)\rangle$ should
show the power behaviour at $|x_{12}|\to\infty$, provided the
spontaneous BRST symmetry breakdown takes place.

\section{Conclusions}
\setcounter{equation}{0}

In this paper we have studied the topological $\sigma$ model
with the black hole metric. Our results for correlation function
$\langle O(x_1),O(x_2)\rangle$ in (6.17) and (6.20) show its
coordinate dependence. This is consistent with the nonzero
result (5.8) for the vacuum energy and ensures the BRST-symmetry
breaking.

The instanton vacuum of the model is equivalent to CG/SG
system. This ensures the IR- finiteness of the physical
observables due to the Debye screening phenomenon in the Coulomb
plasma. The temperature of CG corresponds to $\beta=1$. This
means that the physical content of our theory is very simple:
we deal with free massive fermions. Actually, we have proved
the equivalence of the black hole model to free massive fermions
in the weak coupling limit $r^2 \gg 1$, because we have not
studied possible perturbative corrections
on top of instanton effects.
However, the topological nature of the model suggests that
this holds true to any order in $r^2$ (the dependence on
 $r^2$ come only in combinations $g_I$ and $g_{\bi}$).

Let us now address a question: is the breakdown of the BRST-symmetry
we observed in the black hole model an explicit one or a
spontaneous
one?  One possible answer is that it is explicit and related
to some sort of a holomorphic anomaly, like the one discovered in
the topological gravity \cite{19}. The argument in favour
of this assertion is that our results
for observables depend
 on  $r^2$. Coupling constant  $r^2$ is the
coefficient in front of the Q-exact operator. The dependence
on such coupling constants is interpreted in \cite{19} as a
holomorphic anomaly.

However, it seems more plausible to interpret the breakdown
of the BRST-symmetry here as a spontaneous one. One argument for this
is that the effective $I$ and $\bi$ vertices (3.7) and (2.29)
are Q-closed. This means that the effective action is
Q-invariant and this is the choice of the vacuum state
that breaks down the Q-symmetry. Another argument is the
power behaviour (6.20) for the correlation function (2.22)
at large distances. We interpret it as a propagation of
the goldstino fermion associated with the spontaneous
BRST-symmetry breaking.

Let us stress however, that the appearance of UV divergences
in our results for physical observables shows the
presence of the new conformal
anomaly. Note, that on the level of perturbation theory the model
has conformal anomaly (the model is not a conformal invariant
one) \cite{20},
 but the $\beta$ function associated
with this anomaly does  not contribute to physical observables
\cite{1,2}. The new anomaly is of a non-perturbative
nature and related to the noncompactness of the modular
space of instantons.

{}From the point of view of the SG description  of our model (3.13)
this anomaly is associated with the tachyon operator
$cos2\phi$. The coupling constant $g_q$ in front of this
operator (which is related to couplings $g_I$ and $g_{\bi}$
via (3.15)) is renormalized according to the RG flow of the
SG model. In particular, the fermion mass as defined
in (3.27) is the RG-invariant.

The BRST-symmetry breaking we observe in this paper can have an
interesting string theory application. The topological
version of the  $SL(2,R)/U(1)$ coset model (which is interpreted
in \cite{6} as c=1 string) differs from our model by the
presence of the dilaton term. The dilaton term is a quantum correction
and can not affect drastically the instanton physics. Then the
emergence of instantons could produce dramatic consequences for
the string theory.
Of course, if the conformal invariance of 2D theory is broken it
cannot serve as a string vacuum state any longer. However, if we think of
quantum string theory, we might have to consider these states as well.
This point of view has been taken up in refs.\cite{KE}.
In particular, in
 papers \cite{JE} the RG flow which could occur in
certain black hole $ \sigma$-models if instantons are taken into account
(the model considered in this paper is an example)
is interpreted as a decay of the false string vacuum and related to the
black hole information loss paradox \cite{Haw}.

The author is grateful to N.Dorey and A.Johansen for stimulating
discussions and to the Particle Physics group of the University
 of Wales Swansea where part of this work was done for hospitality.
This work was supported by the
Higher Education Funding Council for
Wales, by the Russian Foundation for Fundamental Studies under Grant
No.93-02-3148 and by Grant No. NOD000 from the Internation Science
Foundation.

\end{document}